\documentclass{pasj00}

\SetRunningHead{Tachihara, Neuh\"auser, \& Fukui}{Molecular clouds in TW Hya}
\Received{}
\Accepted{2009/03/04}
\Published{}

\begin{document}
\title{Search for Remnant Clouds Associated with the TW Hya Association
\thanks{ESO run ID: 071.C-0568}}
\author{Kengo \textsc{Tachihara}
\thanks{Present address: National Astronomical Observatory of Japan, 
2-21-2, Mitaka, Tokyo, 181-8588, Japan; k.tachihara@nao.ac.jp}}
\affil{Department of Earth and Planetary Sciences, Graduate School of Science, 
Kobe University,\\ 1-1 Rokko-dai, Nada-ku, Kobe, 657-8501, Japan}
\email{tatihara@kobe-u.ac.jp}
\author{Ralph \textsc{Neuh\"auser}}
\affil{Astrophysikalisches Institut und Universit\"ats-Sternwarte Jena, 
Schillerg\"a{\ss}chen 2-3, D-07745, Jena, Germany}
\email{rne@astro.uni-jena.de}
\and
\author{Yasuo \textsc{Fukui}}
\affil{Department of Astrophysics, Nagoya University, Chikusa-ku, 
Nagoya, 464-8602, Japan}
\email{fukui@a.phys.nagoya-u.ac.jp}
\maketitle

\begin{abstract}
We report on a search for the parental molecular clouds  of the TW Hya association 
(TWA), using CO emission and Na\emissiontype{I} absorption lines.  
TWA is the nearest young ($\sim 50$~pc; $\sim 10$~Myr) stellar association, 
yet in spite of its youth, there are no detection of any associated natal molecular 
gas, as is the case for other typical young clusters.  
Using infrared maps as a guide, we conducted a CO cloud survey toward a region 
with a dust extinction of $E(B-V) > 0.2$ mag, or $A_{V} > 0.6$ mag.  
CO emission is detected toward three IR dust clouds, and we reject one cloud 
from the TWA, as no interstellar Na absorption was detected from the nearby 
Hipparcos stars, implying that it is too distant to be related.  The other two clouds 
exhibit only faint and small-scale CO emission.  Interstellar Na\emissiontype{I} 
absorptions of Hipparcos targets, HIP 57809, HIP 64837, and HIP 64925 (at 
distances of 133, 81, and 101~pc, respectively) by these couds is also detected.  
We conclude that only a small fraction of the interstellar matter (ISM) toward 
the IR dust cloud is located at distance less than 100 pc, which may be all that 
is left of the remnant clouds of TWA; the remaining remnant cloud having 
dissipated in the last $\sim 1$ Myr.  
Such a short dissipation timescale may be due to an external perturbation or 
kinematic segregation that has a large stellar proper motion relative to the 
natal cloud.  
\end{abstract}

\KeyWords{Open clusters and associations: individual (TW Hya) --- 
Stars: formation --- ISM: bubbles --- ISM: clouds ---  Radio lines: ISM}

\section{Introduction}

\citet{kastner97} examined a few young stars near the classical T Tauri star, 
TW~Hya (which has become known as the TW Hya association), and suggested 
that they are nearby and of a young age.  Later Hipparcos parallaxes measurements 
towards TW Hya and HD 98800 yielded a distance of $\sim 50$~pc, and an age 
of $\sim 10$~Myr, although the cluster age and its dispersion are still unclear 
\citep{tachihara03}.  The stellar population number of TWA has expanded after a 
ROSAT X-ray survey and other followup measurements (e.g., \cite{webb99}), and 
the current population of the association now comprises more than 20 stars 
\citep{zuckerman01,webb01}.  

The proximity of TWA makes it an ideal source for studying faint objects such as: 
extra-solar planets, protoplanetary disks, and young brown dwarfs \citep{ralph00}.  
Notably however, despite its young age, no molecular cloud remnants from the 
formation of TWA have been found.

The ROSAT all sky survey and its follow-up optical observations have been successful 
in finding a large number of T Tauri stars at a range of distances to their nearest 
molecular cloud (e.g., \cite{ralph97}).  At this stage, the process by which a T Tauri star 
becomes separated from its nursery molecular cloud is unclear; one plausible scenario 
postulates that the young stars are formed in relatively small clouds which are rapidly 
dissipated \citep{mizuno98,tachihara01,tachihara05}.  In general however, young stars 
are found to be closely associated with their natal molecular clouds.  For example; 
Taurus, Chamaeleon, and Lupus 3, whose ages are a few, 5, and a few $\times 10$~Myr, 
respectively, are all associated with a rich molecular cloud or envelope.  An older 
example, the Pleiades (125~Myr) on the other hand, is not associated with any 
molecular gas.  

A notable young exception is the $\eta$ Chamaeleontis cluster ($\sim 8$ Myr) at 
$\sim 97$ pc, which has no surrounding molecular gas despite its young age 
\citep{mamajek99, mizuno01}.  \citet{mamajek00} proposed that the $\eta$ Cha and 
TW Hya clusters were formed as parts of Sco-Cen OB association and later dispersed 
with relatively large proper motion.  The proximity of TWA makes it better target for 
exploring this scenario in detail, as a search for remaining faint and small remnant 
clouds can be made more easily.

As the star formation efficiency for typical star forming regions (SFRs) is much less than 
50\%, where the remaining molecular gas is returned to the ISM, the study of the 
dissipation of molecular gas is an important stepping-stone in the overall objective 
of understanding matter cycling in the Galaxy.  
The largest and most extensive molecular cloud survey in our Galaxy is the CfA CO 
survey (e.g., \cite{dame01}).  However, this survey does not cover the distribution 
of the TWA members and it is not of sufficient sensitivity to detect the expectedly 
faint CO of the TWA molecular cloud remnant.

\section{Observations}

\begin{figure*}
\begin{center}
\FigureFile(150mm,120mm){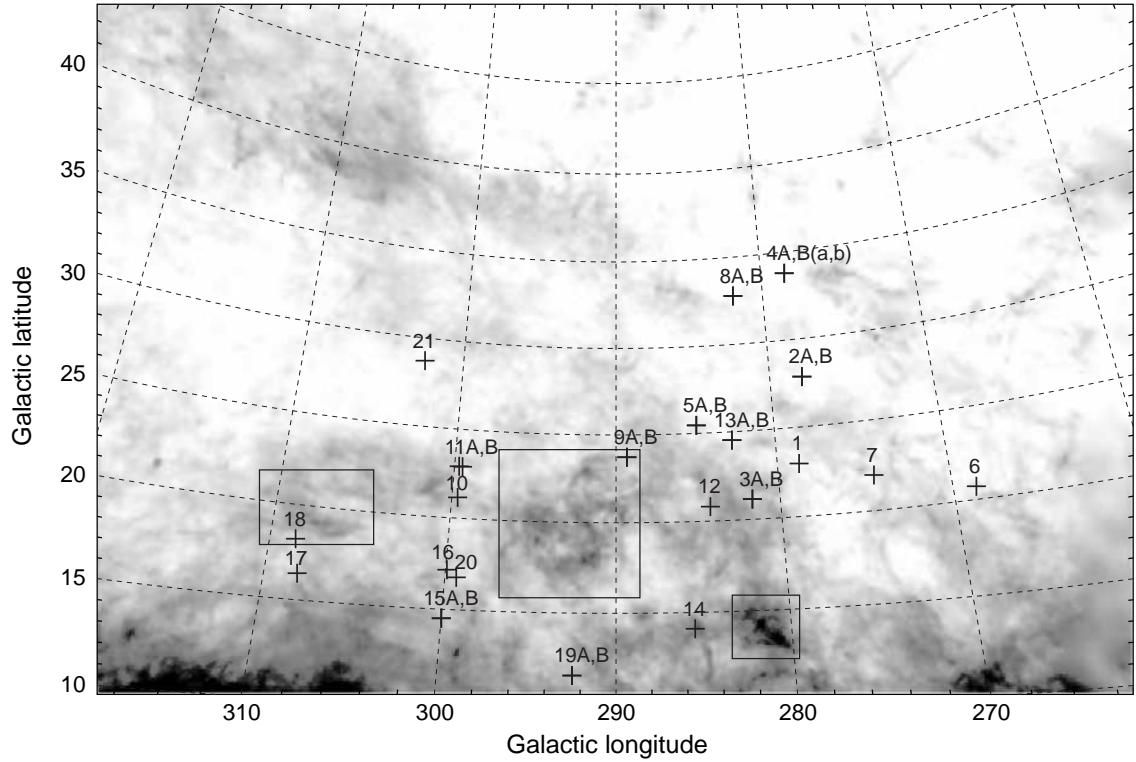}
\end{center}
\caption{Far-IR dust $A_V$ map derived from IRAS 100 $\mu$m and 
COBE/DIRBE 240 $\mu$m surveys \citep{schlegel98}.  The TWA stars 
are shown by crosses with numbers.  Three boxes designate the IR cirrus clouds 
named cloud A, B, and C from right to left. }
\label{distribution}
\end{figure*}

We have made a search for any remnant natal molecular cloud associated with TWA 
using  the CO ($J=$1--0) emission line at 115.2712~GHz and Na\emissiontype{I} doublet 
absorption lines at $\lambda = 5889.95$~\AA\ and 5895.92~\AA.  
TWA subtends an approximate $40^{\circ} \times 30^{\circ}$, and we limit the survey 
size by using the the IRAS far-IR survey data as a guide map; Fig.~\ref{distribution} 
shows a dust extinction map of $E(B-V)$ derived from IRAS 100 $\mu$m and 
COBE/DIRBE 240 $\mu$m, as demonstrated by \citet{schlegel98}.  We note that 
some infrared cirrus clouds are located above the Galactic plane where TWA-member 
stars are distributed.  However, these clouds do not appear to spatially correlate with 
the TWA-member stars.  The typical age of TWA stars is $\sim 20\pm10$ Myr 
\citep{kastner97, jensen98}, and TWA members with proper motion of 1 km s$^{-1}$ 
will travel $\sim 10$ pc ($\sim 11$ deg) since their formation, effectively separating 
them from the natal clouds.  As such, we focus on relatively intense 
and large IR cirrus clouds, not biased onto spatial correlation of a TWA-member.  
We instead selecting IR cirrus clouds whose $E(B-V)$ peaks are greater than 
0.2 mag (corresponding to $A_{V} > 0.6$ mag, assuming the empirical relation 
of $A_{V} = R_{V}\,E(B-V)$ with $R_{V} \simeq 3.1$).  We identify one large and 
intense cloud, and two diffuse and extended clouds above $b > 15\deg$ that 
contain several local peaks.  Hereafter, we refer to these three prominent IR 
cirrus clouds shown in Fig.~\ref{distribution} as cloud A, B, and C.  

\subsection{CO observations}
$^{12}$CO ($J=$1--0) observations were conducted with the 4m NANTEN radio 
telescope at Las Campanas observatory in Chile.  
At 115 GHz, NANTEN has a beam FHWM of 2.6 arcmin, and a velocity resolution 
of $\sim$ 0.1~km~s$^{-1}$.  Regions where the dust $E(B-V) > 0.2$ mag were quickly 
surveyed with a 4-arcmin grid spacing, to a rms sensitivity of $\sim 0.2$~K.  Resulting 
regions containing a possible CO detection were resurveyed to improve the sensitivity 
to $< 0.1$~K.  These observations were made with a frequency-switching technique, 
using a throw interval of 13 -- 20 MHz depending on atmospheric condition.  
A position-switching technique was used instead for clouds with a $V_{\rm LSR}$ 
close to the telluric CO line. In total, 764 positions were observed.

\subsection{Na\emissiontype{I} obsercations}
We  check the physical association of the target molecular clouds with TWA through 
a distance comparison: 
TWA-member stars have a distance range of 30 to 120~pc \citep{frink01}.  Therefore, 
any background stars in the same line of sight of the clouds, and with a distance of 
$d \geq 120$~pc should be measurably extincted by the candidate natal clouds.  
We selected 21 Hipparcos stars with a parallactic distances $d < 150$~pc, located 
in and around high dust column density regions of cloud A--C.  These were observed 
with the European Southern Observatory (ESO) 2.2m telescope using the FEROS 
(Fiber-fed Extended Range Spectrograph).  The spectrum range of FEROS spans 
3600 to 9200~\AA, and has a resolving power or $R = 48000$ (2 pixels resolution).  
The standard MIDAS pipeline was used for bias subtraction, flat-fielding, scattered 
light removal, sky background subtraction and wavelength calibration.  

\section{Results}

\subsection{CO survey}

\begin{figure}
\begin{center}
\FigureFile(80mm,100mm){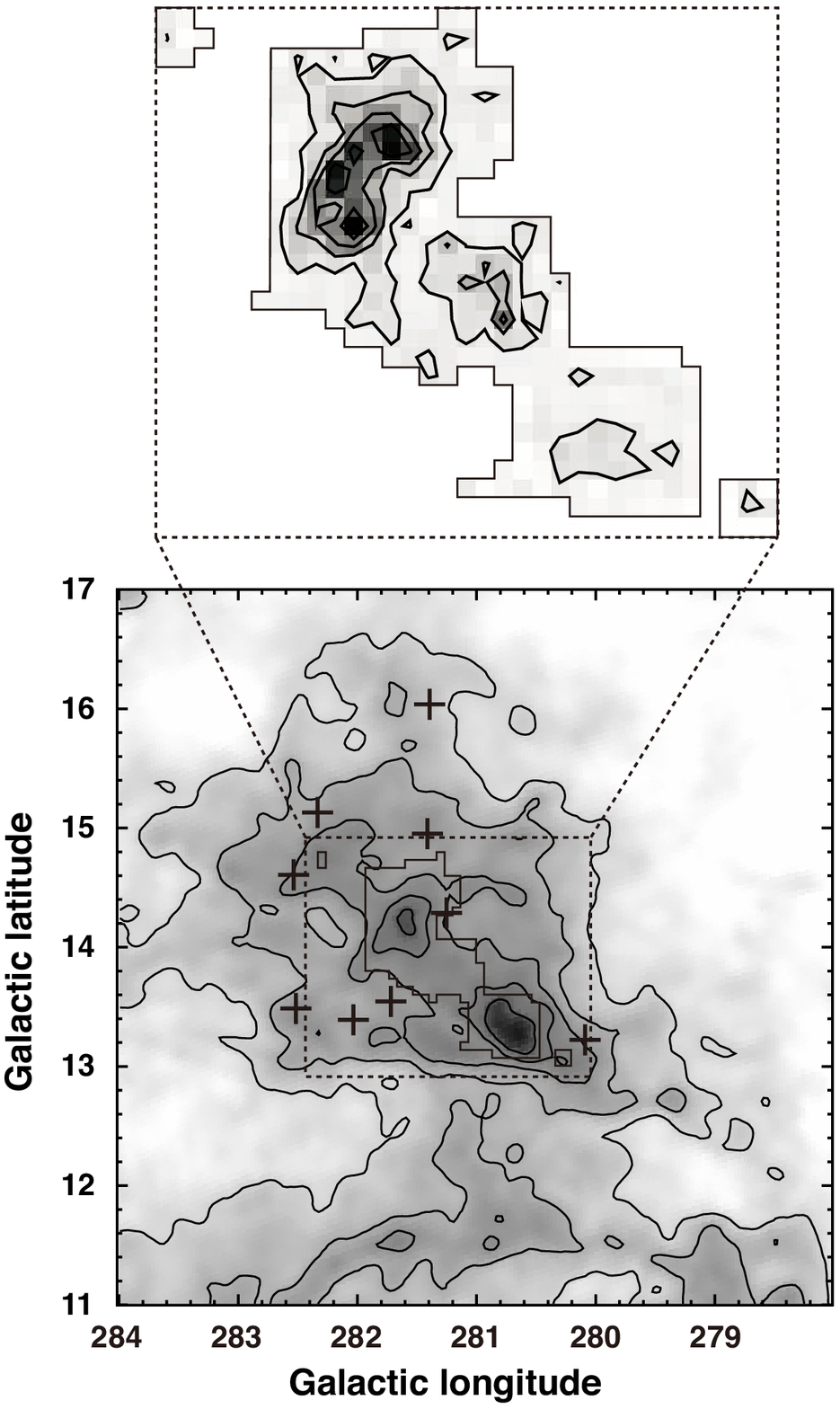}
\end{center}
\caption{Bottom: Dust $E(B-V)$ map of cloud A.  Contour intervals are 0.15+0.05 
mag and the crosses indicate the positions of the observed Hipparcos stars.  
Top: CO integrated intensity map of cloud A, over the region indicated in the 
dust map below.  The solid lines indicate the limits of the CO surveyed region.  
Contour intervals are 1.7+1.7~K~km~s$^{-1}$.}
\label{cloudA}
\end{figure}

The Integrated CO intensity map and the $E(B-V)$ map for cloud A is shown in 
Fig~\ref{cloudA}, where we see relatively strong CO emission and its intensity 
distribution is generally similar to that of dust with $A_{V}> 0.84$.  Notably however, 
the strong dust peak at $l \sim \timeform{280D.6},\ b \sim \timeform{13D.5}$ 
is weakly represented in the CO map.  Conversely, the second strongest dust 
peas at $l \sim \timeform{281D.5},\ b \sim \timeform{14D.5}$ is strongly represented 
in the CO map.  In addition, we find another CO peak located between the two dust 
peaks.  The peak CO integrated emission is 9.6 K km s$^{-1}$, corresponding to 
a molecular hydrogen column density of $1.5 \times 10^{21}$ cm$^{-2}$ 
\citep{hunter97}.  This is roughly consistent with that derived from the dust 
$A_V$ ($\sim 1.1$ mag), assuming all the hydrogen is in the molecular form, 
i.e., $N({\rm H_2}) / A_{V} = 10^{21}$ cm$^{-2}$ mag$^{-1}$ \citep{bohlin78}.  

\begin{figure*}
\begin{center}
\FigureFile(120mm,200mm){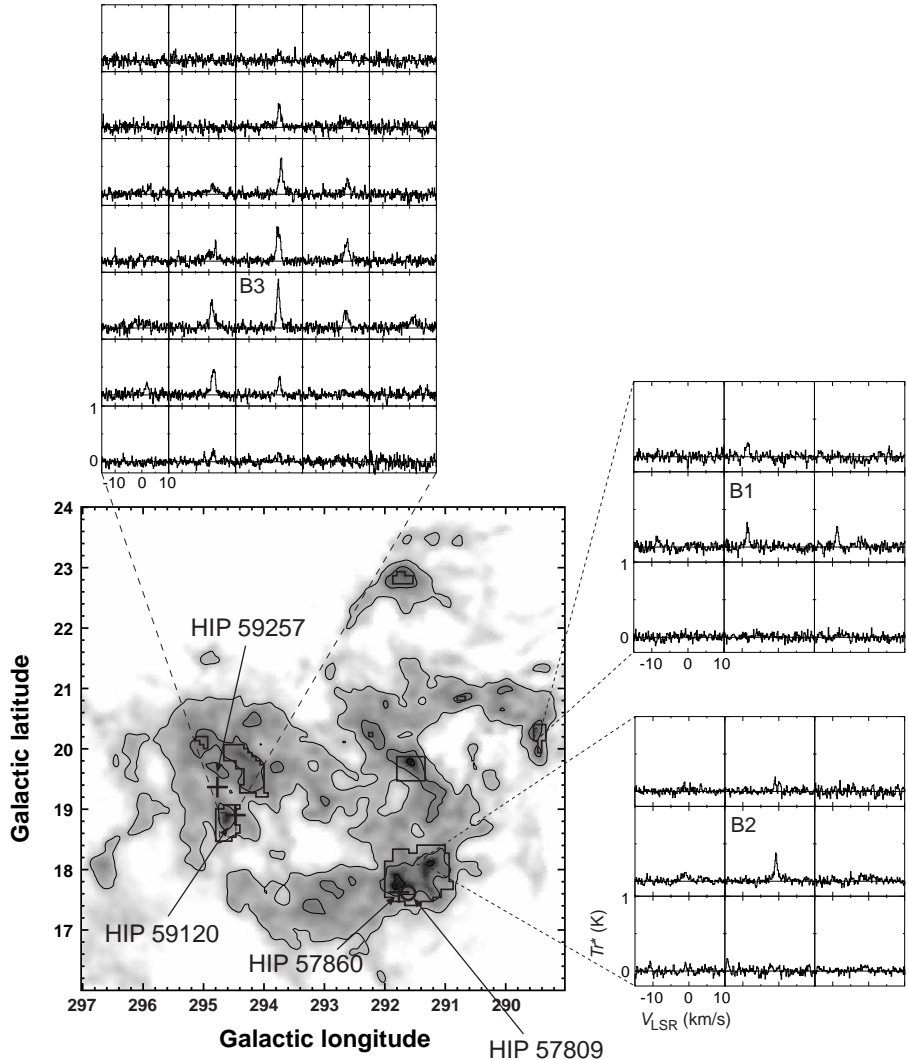}
\end{center}
\caption{Main: The dust $E(B-V)$ map of cloud B.  Contour intervals are 0.15+0.05 
mag and the crosses indicated the position of the observed Hipparcos stars. Aside: 
Spectral maps for the three detected CO peaks.  CO peak positions listed in 
Table \ref{COpeaks} are labeled.  The circle shows the position of HIP 57809 
that shows interstellar Na absorption, while crosses are other HIP stars without Na 
detections.}
\label{cloudB}
\end{figure*}

\begin{figure*}
\begin{center}
\FigureFile(120mm,100mm){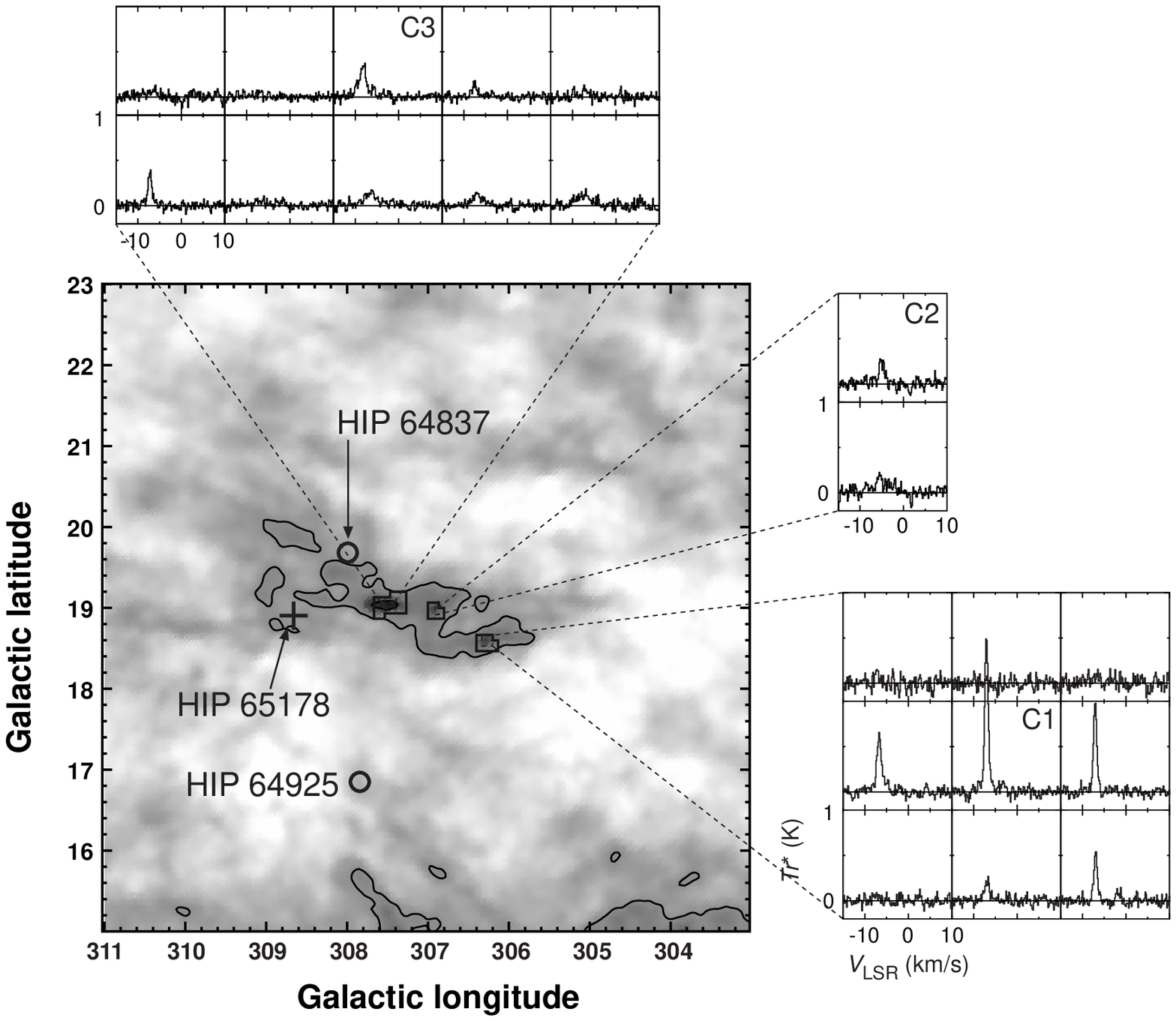}
\end{center}
\caption{Main: The dust $E(B-V)$ map of cloud C.  Contour intervals are 0.15+0.05 
mag and the crosses indicated the position of the observed Hipparcos stars. Aside: 
Spectral maps for the three detected CO peaks.  CO peak positions listed in 
Table \ref{COpeaks} are labeled.  The circles show the positions of HIP 64837 
and HIP 64925 that show interstellar Na absorption, while the crosse is HIP 65178 
without Na detection.}
\label{cloudC}
\end{figure*}

In contrast to cloud A, CO emission is detected from only small and weak regions 
in clouds B and C (Figs.~\ref{cloudB} and \ref{cloudC}).  In total, emission with 
$W{\rm (CO)} > 0.3$ K km s$^{-1}$ was measured at only 37 positions from total 
of 436 points, as shown in Figs.~\ref{cloudB} and \ref{cloudC}.  
Table \ref{COpeaks} lists the properties of the brightest three $W{\rm (CO)}$ peaks 
each from clouds B and C.  The dust maps for these two clouds shows a complex 
morphology, with several column density peaks having $A_{V} > 0.6$ mag.  

CO emissions are detected only toward the three of many peaks of cloud B.  
In contract, CO was detected towards all the three dust peaks of cloud C.  
The correlation between dust $A_{V}$ and CO intensity is generally weak.  
The CO spectra towards both clouds B and C, shown in Figs.~\ref{cloudB} and 
\ref{cloudC}, are generally complex and have a number of velocity components.  
The systemic velocity of the CO emission components range between 
$-9$ to $+1$~km~s$^{-1}$ in cloud B, and from $-8$ to $-5$~km~s$^{-1}$ 
in cloud C.  These velocities are slightly smaller than the mean $V_{\rm LSR}$ 
of the TWA stars, at +2.8~km~s$^{-1}$ \citep{torres03}.  Additionally worthy of note, 
is that the $V_{\rm LSR}$ of the CO emission peaks are more dispersed than 
velocity dispersion of the TWA-member stars (1.9~km~s$^{-1}$ as a standard deviation).

\begin{table*}
\caption{Properties of CO peaks}\label{COpeaks}
\begin{center}
\begin{tabular}{cccccccccc}
\hline
name & \multicolumn{2}{c}{Galactic} & R.A. & Decl.& $W$(CO) & 
peak $T_{\rm r}$\footnotemark[$*$] & $V_{\rm LSR}$\footnotemark[$*$] & 
$\Delta V$\footnotemark[$*$] & $N{\rm (H_2)}$\footnotemark[$\dagger$] \\
& longitude & latitude & (J2000.0) & (J2000.0) & (K km s$^{-1}$) & (K) & (km s$^{-1}$) 
& (km s$^{-1}$) & ($10^{20}$ cm$^{-2}$)\\
\hline
B1 & 289.4667 & 20.2667 & 11 44 18.3 & $-$40 50 35 & 0.32 & 0.27 & $-8.5$ & 0.9 & 0.5 \\
B2 & 291.4000 & 18.1333 & 11 50 49.8 & $-$43 22 38 & 0.57 & 0.32 & $-0.8$ & 1.1 & 0.9 \\
B3 & 294.6000 & 18.6667 & 12 07 48.1 & $-$43 29 53 & 0.93 & 0.75 & $\phantom{+}1.0$ & 1.2 & 1.5 \\
C1 & 306.2667 & 18.6000 & 13 09 04.0 & $-$44 09 27 & 2.02 & 1.71 & $-7.0$ & 0.9 & 3.2 \\
C2 & 306.8667 & 19.0000 & 13 12 02.3 & $-$43 42 51 & 0.51 & 0.26 & $-5.0$ & 1.5 & 0.8 \\
C3 & 307.4667 & 19.0667 & 13 15 07.7 & $-$43 35 47 & 0.90 & 0.31 & $-8.3$ & 2.4 & 1.4 \\
\hline
\multicolumn{10}{@{}l@{}}{\hbox to 0pt{\parbox{180mm}{\footnotesize
\footnotemark[$*$]{Derived by fitting to the gaussian profiles.}
\par\noindent
\footnotemark[$\dagger$]{Assuming the conversion factor $N({\rm H_2})/W({\rm CO})$ 
of $1.56 \times 10^{20}$ cm$^{-2}$ (K km s$^{-1})^{-1}$ \citep{hunter97}.}
}\hss}}
\end{tabular}
\end{center}
\end{table*}

\subsection{Optical spectroscopy}

\begin{figure*}
\begin{center}
\FigureFile(180mm,50mm){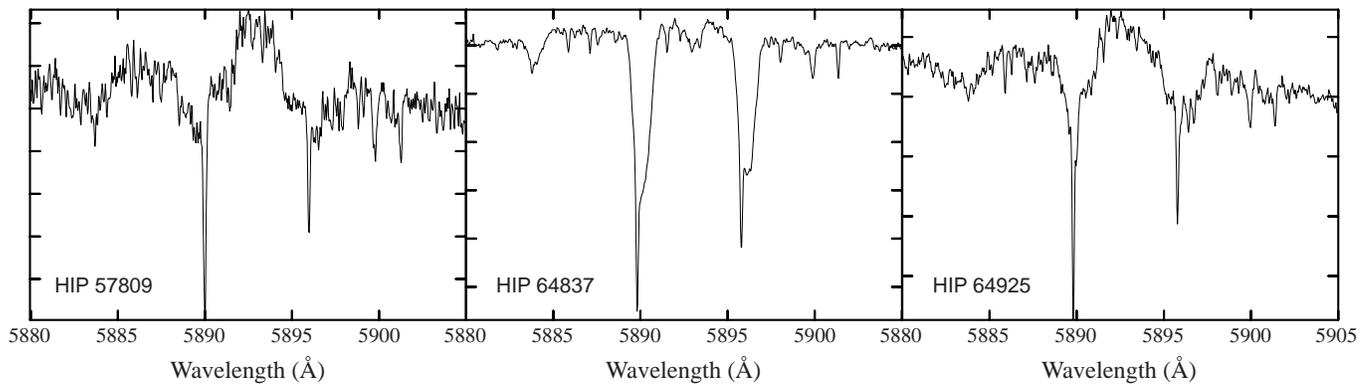}
\end{center}
\caption{Optical spectra of HIP 57809, HIP 64837, and HIP 64925.  The ordinate 
is the normalized relative intensity.  They all show composite Na absorption lines 
that consist of wide stellar components 
and narrow interstellar ones.  The intensities are in arbitrary scales.  
}
\label{spectra}
\end{figure*}

To help distinguish the blended Na\emissiontype{I} stellar emission lines and interstellar 
absorption lines, we have defined an absorption detection according to the following 
three criteria: (a) Both of the Na doublet lines have two (partially) resolved components; 
(b) the interstellar lines are significantly narrower than the stellar ones; (c) the star is not 
a spectroscopic binary, according to many other absorption lines in the spectrum.  
Applying these criteria result in the detection of interstellar Na\emissiontype{I} D 
absorption of three targets; HIP 57809, HIP 64837, and HIP 64925 (Fig.~\ref{spectra}).  
Targets HIP 57809 and HIP 64837 are towards clouds B and C, respectively, and 
HIP 64925 is located $\sim 2$ degrees away from cloud C.  The parallactic distances 
for HIP 57809, HIP  64837 and HIP 64925 are 133, 81, and 101~pc, respectively.  
We do not detect absorption for any targets towards cloud A.  

The equivalent width, $W({\rm Na})$ and $V_{\rm LSR}$ (calculated from the central 
wavelength) of the interstellar lines were estimated by fitting a double-component 
gaussian.  To avoid confusion with atmospheric lines, we estimate $W({\rm Na})$ 
from the D2 line only.  After de-blended from the stellar absorption lines we 
estimate a $W({\rm Na})$ of 0.01, 0.04, and 0.01~\AA\ and $V_{\rm LSR}$ of 
13.6, $-1.4$, and $-2.5$~km~s$^{-1}$ for HIP 57809, HIP 64837, and HIP 64925, 
respectively.  The optical spectral resolution relative to the radial velocity is 6.1 km 
s$^{-1}$, which is poorer than that obtained from the radio observations.  

\section{Discussion}

\subsection{Distance to the clouds}

We use the Na absorption lines to constrain the distance to the three clouds. 
Nine Hipparcos stars with spectroscopic data are in the direction of cloud A, 
however none of these show evidence for absorption, thus the distance of the 
molecular cloud is likely to be greater than that of the nine stars, specifically, 
$D_{\rm A} > 144$~pc.  
The mean distance of TWA-members is much less than this, and so we conclude 
that cloud A is not associated with the TWA. The total cloud mass is derived is 
estimated by
$\frac{M}{\MO} = 67.5 \left( \frac{d}{\rm 150~pc} \right) ^{2}$.

Determining the distances to clouds B and C is complicated by their fragmentary 
morphology.  Four Hipparcos stars were observed towards cloud B and only one 
of them, HIP 57809 ($d=133$~pc), shows interstellar absorption.  A nearby 
star ($d=92$~pc), HIP 57860, does not show evidence for absorption.  
We can thus constrain the distance to cloud B (and more specifically the dust 
peak at $l \sim \timeform{291D.8},\ b \sim \timeform{17D.7}$ east of the CO peak B2) 
to $92~{\rm pc} < D_{\rm B2\,east} < 133~{\rm pc}$.  
The remaining two stars (HIP 59120 and HIP 59257, at distances of 108 and 76~pc, 
respectively) do not show interstellar absorption, and so we can further constrain 
the distance estimate for the peak at $l \sim \timeform{294D.5},\ b \sim 
\timeform{19}$ to $D_{\rm B3} > 108$~pc.  It is worth mentioning that the velocity of the 
interstellar absorption toward HIP 57809 ($V_{\rm LSR} = +13.6$~km~s$^{-1}$) 
is which is significantly larger than that of CO emission at the nearby peak B2 
($V_{\rm LSR} = -0.8$~km~s$^{-1}$), suggesting two layers of ISM in the line of sight.  
Large difference in $V_{\rm LSR}$ of the three CO peaks (from $-8.5$ to $+1.0$ 
km s$^{-1}$) also imply multiple components.  

Cloud C shows similar level of complexity:  Two Hipparcos stars are located 
close to this cloud (HIP 64837, $d=81$~pc and HIP 65178, $d=128$~pc) 
and only HIP 64837 shows evidence for absorption.  A third star, HIP 64925 
($d=101$~pc), also evidence for the absorption, however this star is somewhat 
more distance from the cloud (see Fig.~\ref{cloudC}).  Together, this suggests that 
the cloud is at a distance of $D_{\rm C} < 81$~pc, although the nearby star 
HIP 65178 at $d=128$~pc does not show absorption possibly due to 
fragmentary small scale cloud morphology.  
The velocities of the Na absorption lines ($V_{\rm LSR} = -1.4$ km s$^{-1}$) 
and CO emission lines ($V_{\rm LSR} = -8.3$ km s$^{-1}$) are again significantly 
discrepant, although they are within the errors of the absorption resolution.  
We find, that the absorption and emission spectra presented here suggest a 
complex  multi-component spatial and kinematic distribution of clouds B and C.

\subsection{Natal clouds of the TW Hya association}

Considering the region encompassing TWA members that is not contaminated 
from the Galactic plane dust emission, $\timeform{267D} < l < \timeform{310D}$ and 
$\timeform{16D} < b < \timeform{35D}$, we find that only $\sim 3\%$ or 23 deg$^2$ 
shows a dust $E(B-V) > 0.15$ mag.  The survey data presented in this report samples 
1.94 deg$^2$ (excluding cloud A), showing CO in emission at only 8.5\% of the 
observed positions.  We can extrapolate this detection ratio to regions where  
$E(B-V) > 0.15$ mag, and estimate that 1.9 deg$^2$ will show CO in emission.  
Assuming a typical CO intensity of 0.5 K km s$^{-1}$ and a molecular hydrogen 
column density of $8 \times 10^{19}$ cm$^{-2}$, we futher estimate that the 
total molecular mass throughout the region defined above is $\lesssim 3~\MO$.

We can also estimate the mass of original pre-stellar molecular cloud; assuming 
a total current stella mass of $\sim 16~{\MO}$ \citep{tachihara03} and a star-formation 
efficiency of 3\% \citep{tachihara02}, yields a pre-stellar molecular cloud mass of 
540 \MO.  As such, less than 1\% of the original cloud mass remains since the most 
recent TWA star formation, approximately 1 Myr ago.  This is a considerably short 
dissipation timescale, in comparison to other nearby SFRs; 
\citet{tachihara01} noted that most T Tauri stars younger than 5 Myr found in 
Lupus and Chamaeleon, are within a few pc of the associated molecular clouds.  
They also note that small isolated clouds may also have been dissipated rapidly by 
nearby OB associations.  

The present velocity dispersions of the candidate TWA natal molecular clouds are 
too high to be bound, suggesting that they may have been recently disturbed by 
some external event.  In the following, we present scenarios that may result in the 
observed and anomalously high velocity dispersions of the candidate TWA molecular 
clouds.

\subsubsection{Scenario I}
Careful analysis of kinematic data suggest that both the TWA and $\eta$ Cha cluster 
formed from the Sco-Cen giant molecular cloud (GMC) at a distance of $\sim 150$ pc 
\citep{mamajek00}.  Over the last $\sim 10$ Myr, they have been moving away from the 
Sco-Cen complex and are presently separated from the Lower Centaurus Crux (LCC) 
subgroup by 50-100 pc.  The Sco-Cen GMC was later evacuated by stellar winds 
and/or SN explosion forming the giant ``Loop I bubble" H\emissiontype{I} shell 
\citep{degeus92}.  The progenitor of the $\eta$ Cha cluster may have been dispersed 
by the same SN responsible for forming the Loop I bubble.  

\subsubsection{Scenario II}
TWA is within 110 pc of the sun, and is inside a low-density plasma region that fills the 
local bubble, observed as isotropic soft X-ray radiation  (e.g. \cite{tanaka77}).  The Pleiades 
moving group is a nearby OB association that may have been the source of a SN explosion 
in the last $\sim 1$ Myr, or of 3 SNe explosions in the last 5 Myr \citep{berghoefer02}.  
The shock wave from these events may be responsible enhancing the dissipation of the 
TWA natal molecular clouds, however the Pleiades moving group is large, with a scale 
of $\sim 100$ pc, and as such the relative proximity of the SN events to TWA is unclear, 
furthermore, no candidate SNe of such an explosion has yet been identified. 

More accurate measurements of the proper motions, distances and radial velocities 
of both the TWA-member stars and candidate remnant molecular clouds, are crucial 
to refine and discriminate between these scenarios, as would search for any remnants 
of any SNs in the Pleiades moving group.

\section{Summary}

We have conducted a search for remnant clouds associated with the formation of 
TW Hya association in both CO emission, and Na absorption.  
We find that CO emission is detected towards three peaks of IR brightness, clouds 
A, B and C. Cloud A is relatively bright in CO emission and shows strong emission 
in IR, whereas clouds B and C are weaker by more than a factor of 3, and show a 
much more fragmentary distribution of CO emission.  We reject the association of 
cloud A to the TWA, on the basis of a lack of a detection of Na absorption lines which 
indicate that it is much more distant than the TWA.  
Na absorption of Hipparcos targets help to constrain the distances of clouds B and C 
as that only a small fraction of the ISM has a distance less than 100 pc.  
Using a statistical approach, we estimate the total amount of molecular gas in this 
region is less than a few solar masses, and conclude that most of the natal molecular 
clouds associated with TWA has already dissipated with an unusually fast 
dissipation time scale, enhanced perhaps, by an external and energetic disturbance, 
or kinetic segregation with proper motion of the member stars.  

\bigskip

We are grateful to all the staff of the European Southern Observatory for their great 
hospitality.  We thank Erik Muller for reading through the text.  
KT thanks financial support from JSPS (Japanese Society for the Promotion 
of Science).  This work was supported by The 21st Century COE Program: 
Origin and Evolution of Planetary System of MEXT of Japan.

\end{document}